\documentclass[preprint,showpacs,preprintnumbers,amsmath,amssymb]{revtex4-1}

\usepackage{color} 
\usepackage{graphicx} 

\usepackage{dcolumn}
\usepackage{bm}

  \newcommand{\nn}{\nonumber}

\begin{document}


\title{Instability of quantum fluctuation around classical nonlinear massive wave solution in Higgs potential and particle creations}

\author{Yoshio Kitadono}
 \email{kitadono@ncut.edu.tw}
 \affiliation{ 
 Liberal Education Center,
 National Chin-Yi University of Technology,\\
 No.57, Sec.2, Zhongshan Rd., Taiping Dist., Taichung 411030, Taiwan}

\author{Tomohiro Inagaki}
\email{inagaki@hiroshima-u.ac.jp}
\affiliation{ 
	Information Media Center and Core of Research for Energetic Universe,
	Hiroshima University,\\
	No.1-3-2, Kagamiyama, Higashi-Hiroshima, Hiroshima 739-8521, Japan}

\date{\today}

\begin{abstract}
We discuss the dynamics of the quantum fluctuation around the nonlinear massive wave solution in the Higgs potential. In particular, we analyze the stability and instability of the mode function. Using  the stability condition for Hill's equation, we obtain the instability region of the mode function in the quantum fluctuation as the function of the parameters in the mode equation. We show that the two types of the instabilities of the system in the particle number can be understood by the Floquet's exponents in the instability parameter region. The analysis will be useful to understand the dynamics of the quantum field theory around the nonlinear massive wave solution when the classical background slightly deviates from the constant background during the middle stage of the phase transition.     
\end{abstract}

\maketitle

\section{Introduction}
The masses for the elementary particles are prohibited by the gauge symmetry. 
In the standard model (SM) of the elementary particle physics, the masses are generated through so-called the Higgs mechanism \cite{Higgs1,Higgs2,Higgs3}, namely, the masses of the particles are given by the product of the vacuum expectation value (VEV) of the Higgs potential and couplings between the Higgs field to each particles. The Higgs mechanism was experimentally tested at the Large Hadron Collider (LHC) and the Higgs boson predicted by the mechanism was observed at LHC by ATLAS collaboration \cite{Higgs.ATLAS} and CMS collaboration \cite{Higgs.CMS}. The spin and parity of the discovered boson are consistent with the SM prediction \cite{Higgs.spin.ATLAS,Higgs.spin.CMS}. 

The Higgs potential is given by the quartic potential of the Higgs field with the negative mass term, and thus equation of the motion (EOM) is expressed by the nonlinear differential equation. The nonlinear nature not only appear in the Higgs dynamics, but also appears in Yang-Mills theory and other theories. Hence the various nonlinear effects have been studied intensively, such as some analytic results for large order perturbation theory \cite{Bender.Wu.1969}, the relation between the quantized theory around the nontrivial solution and non-perturbative effect \cite{Dashen.1974}, and the effect on the Borel summation \cite{Graffi.1980}. 

The nonlinear solutions in Yang-Mills theory were studied intensively in 1970's, as Coleman suggested the existence of the nonlinear plane wave solution \cite{Coleman.1977}. For instance, the nonlinear plane wave solution in $\mbox{SU}(2)$ Yang-Mills gauge theory was studied \cite{Treat.1971}, the reduction method from gauge theory to scalar theory was found \cite{Corrigan.1977,Oh.1979.1985},  the solution described by Jacobi's $cn$ function was discussed \cite{Baseyan.1979,Matinyan.1981}, the method based on elliptic functions for scalar theory was investigated comprehensively \cite{Actor.1979}, and the solution expressed by Weierstrass elliptic function in $\mbox{SU}(3)$ gauge theory was reported \cite{Tsapalis.2016}. 

In phenomenological aspect, the relation between the infrared gluon in quantum field theory and elliptic function was discussed \cite{Frasca.2008.2009}, the nonlinear solution was used to estimate the Higgs mass \cite{Achilleos.2012}, and the three modes in the classical nonlinear massive solutions described by the Jacobi's $1/dn$-, $dn$-, and $cn$-type solutions in the Higgs potential were discussed \cite{KI.PRD.2016, KI.PLB.2024}. In these analysis, we found that the quantum field theory around the nonlinear massive wave solution with the special limit smoothly reduces to the quantum field theory around VEV. Therefore, it will be worthy developing the formalism of the quantum field theory around the nonlinear massive wave. Especially, it is important to study the particle creation process during the intermediate step of the phase transition during the environment with slightly excited classical background described by the Jacobi's $1/dn$, $dn$, and $cn$ functions. 

The particle creation process through the time dependent background is important to understand the nature of the universe \cite{Ford.2023} both in the context of the theoretical particle physics and cosmology. In particular, the particle creation is related to the instability of the quantum mode \cite{Kofman.1994.1997}. If the solution of the mode equation in the quantum fluctuation around the time-dependent background shows the exponential growth, the particle production in the system is enhanced. To study the stability/instability of the quantum fluctuation around the nonlinear massive wave solution, we must determine the stable/unstable parameter space in the theory. If the oscillation is bounded near the stable point near minimum of the potential, one can use the well-known stability analysis based on the Matheiu equation \cite{Abramowitz, Gradshteyn}. However, the nonlinear massive wave solutions described by the Jacobi's elliptic functions are not simple trigonometric functions and thus we must extend the stability/instability analysis.

In this article, we discuss how to find out stable/unstable parameter region for the parameters of the quantum mode around the nonlinear massive classical wave solution. We will use so-called Hill's equation which is a generalization of the Mathieu equation. The Mathieu equation is described by the differential equation $y''(t)+(a+b\cos(2t))y(t)=0$, while the Hill's equation is expressed by $y''(t)+a(t)y(t)=0$ with an arbitrary periodic function $a(t)$. The periodic function $a(t)$ in our system corresponds to the square of the classical solution described by the Jacobi's $1/dn$, $dn$, and $cn$ functions. According to the Floquet's theorem for Hill's equation, the periodicity of the system constrains many aspects of the solution. We fully apply this method to our analysis on the mode function in the quantum fluctuation and obtain the unstable parameter region and the solution of the Hill's equation. As the consequence of the instability of the mode equation, we discuss the time dependent particle production process. 

It is noteworthy that we ignore the effect of expansion of the universe and thus the amplitude of the nonlinear massive wave does not suppress in this analysis. However we assume that this solution only appears during the intermediate step of the real universe and thus it does not affect any data at the LHC, because the amplitude of the nonlinear wave has already diminished completely.

We briefly review the classical nature of the nonlinear massive wave solution in the Higgs potential with a special attention to the three modes of the oscillation in Sec.~\ref{sec2}. We will introduce our formalism of the quantization and show unstable parameter region which shows the spinodal instability in the mode equation. In Sec.~\ref{sec3}, we will apply the general theory of the Hill's equation to our system. As the results, we obtain the unstable region of our theory and the particle production with some unstable parameters. We discuss the possible implications of the relation between the instability band and particle productions in Sec.~\ref{sec4}. The Sec.~\ref{sec5} is devoted to the conclusion of our analysis.

\section{Formalism: Elliptic oscillating classical solution \label{sec2}}
\subsection{Nonlinear massive wave solution}
First, we consider the Higgs sector in the standard model:
\begin{eqnarray}
\mathcal{L} &=&
 (D_{\mu}\Phi)^{*}(D^{\mu}\Phi) 
 - V(\Phi^{*}\Phi) 
 -\lambda_d \bar{Q}_L\Phi d_R 
 -\lambda_u \bar{Q}_L i\sigma^2\Phi^{*}u_R + \mbox{c.c.},\nn\\
  V(\Phi^{*}\Phi) &=& - \mu^2 \Phi^{*}\Phi
             + \lambda (\Phi^{*}\Phi)^2,
\end{eqnarray}
where $D_{\mu}$ stands for the covariant derivative including $\mbox{SU}_{L}(2)$ and $\mbox{U}_Y(1)$ gauge fields, $\Phi$ is the complex $\mbox{SU}_L(2)$ doublet,  $V(\Phi)$ represents the Higgs potential with the negative mass term $-\mu^2$ and the Higgs self coupling $\lambda$, $\lambda_{u(d)}$ is the Yukawa coupling between the Higgs field to the left-handed-up(down)-quark and to the right-handed-field, $\sigma^2$ is the Pauli matrix, and $\mbox{c.c.}$ stands for the complex conjugate. 

Taking the unitarity gauge which only remains the physical degree of the freedom,
the $\mbox{SU}_{L}(2)$ Higgs doublet $\Phi$ reduces to $\Phi(x) =(0,\phi(x))/\sqrt{2})^{t}$ and then we obtain
\begin{eqnarray}
\mathcal{L}
= \frac{1}{2}(\partial \phi)^2 - V(\phi) + \frac{g^2}{4}W^{+\mu}W^{-}_{\mu}\phi^2 + \frac{g^2 + g^{\prime 2}}{8}Z^{\mu}Z_{\mu}\phi^2 - y_f \bar{\psi}\psi\phi,
\end{eqnarray}
where the Higgs potential $ V(\phi) = -\mu^2\phi^2/2 + \lambda\phi^4/4$, and $g$ and $g^{\prime}$ represent $\mbox{SU}_L(2)$ and $U_Y(1)$ gauge couplings, $W^{\pm}$ and $Z$ stand for the electroweak gauge bosons, and $y_f$ is the scaled Yukawa coupling $(y_f=\lambda_f/\sqrt{2})$ for a fermion $\psi$, respectively.
The above potential has the vacuum expectation value (VEV) at $v \equiv \sqrt{\mu^2/\lambda}$ and the Higgs field $h(x)$ is introduced as the excitation from the VEV though the relation, $\phi(x)=v+h(x)$. As the consequence of the spontaneous symmetry breaking, the Higgs field and other fields acquire the masses.  In this analysis, we ignore the couplings between the Higgs field to other fields for simplicity, because the analysis for the stability/instability for the quantum mode around the classical field is complicated and thus we must analyze the simplest case.

The EOM for the neutral Higgs field $\phi$ is given in
\begin{eqnarray}
	\partial^2 \phi - \mu^2 \phi + \lambda \phi^3 = 0,
\end{eqnarray}
and we can write the nonlinear massive wave solution of the above EOM into the form:
\begin{eqnarray}
	\phi_{\mathrm{cl}}(x) 
= \left\{
\begin{matrix}
 \frac{\phi_0}{dn(p_1\cdot x, k_1)} \hspace{1cm}(0<\tilde{\phi}_0 \le 1),\\
 \phi_0 dn(p_2 \cdot x, k_2) \hspace{1cm}(1 < \tilde{\phi}_0 < \sqrt{2}), \\
 \phi_0 cn(p_3 \cdot x, k_3) \hspace{1cm}(\sqrt{2} < \tilde{\phi}_0),
\end{matrix}
\right.
\end{eqnarray} 
where $\tilde{\phi}_0 \equiv \phi_0/v$ stands for the dimensionless field value normalized by VEV for a given initial field value $\phi_0$, the elliptic modulus $k_{1}$, $k_{2}$ and $k_{3}$, and the square of the four momenta in the classical solution denoted by $p^2_{1}$, $p^2_{2}$ and $p^2_{3}$ are obtained as the function of $\tilde{\phi}_0$ in Ref.~\cite{KI.PLB.2024}. We simply denote $p^2\equiv m^2_{\mathrm cl}$ for each three modes in this article.

\subsection{Quantization around time dependent elliptic oscillation}
We combine the classical nonlinear massive solution with the formalism of the quantization in Ref.~\cite{Herring.2024} to study the dynamics of the quantum fluctuation which couples to the time-dependent-classical solution. First, we introduce the quantum fluctuation $h(x)$ around the classical solution $\phi_{\mathrm cl}(x)$ by
\begin{eqnarray}
	\phi(x) = \phi_{\mathrm{cl}}(x) + \sqrt{\hbar}h(x),
\end{eqnarray}
and take the rest frame of the four momentum of the classical field, $p_{i}^{\mu}=(m_{\textrm{cl}}, \vec{0})$ for each parameter region ($i=1,2$, and $3$), so that the classical solution only depends on time. This choice is always guaranteed, because the squared momentum of the classical solution is massive. Then the expectation value of the field $\phi(x)$ for the coherent state $|\Omega\rangle$ is given by
\begin{eqnarray}
	\langle \Omega|\phi(x)|\Omega\rangle = \phi_{\textrm{cl}}(t),
\end{eqnarray} 
where the expectation value depends on time. The EOM for the quantum fluctuation $h$ up to $O(\hbar)$ is given in
\begin{eqnarray}
	\ddot{\phi}_{\mathrm{cl}}(t) + V'(\phi_{\mathrm{cl}}(t)) + \partial^2{h} + V''(\phi_{\mathrm{cl}}(t))h = 0,
\end{eqnarray}
where $\dot{\phi}$ stands for the derivative with respect to the time variable $t$ and $V'(\phi)$ stands for the derivative with respect to the field $\phi$, and we take $\hbar=1$. 

Next we introduce the quantization of $h$ field and its conjugate momentum $\pi_{h}$:
\begin{eqnarray}
 h(x) = \sqrt{\frac{1}{V}}\sum_{\vec{p}}
 \left[ a_{\vec{p}}g_{p}(t)e^{i\vec{p}\cdot \vec{x}} + a^{\dagger}_{\vec{p}}g^{*}_{p}(t)e^{-i\vec{p}\cdot \vec{x}} 
 \right], \\
 \pi_h(x) = \sqrt{\frac{1}{V}}\sum_{\vec{p}}
\left[ a_{\vec{p}}\dot{g}_{p}(t)e^{i\vec{p}\cdot \vec{x}} + a^{\dagger}_{\vec{p}}\dot{g}^{*}_{p}(t)e^{-i\vec{p}\cdot \vec{x}} 
\right],
\end{eqnarray}
where $V$ is the spatial volume of the system and the mode function $g_{p}(t)$ is the solution of the mode equation: 
\begin{eqnarray}
 0 &=& \frac{d^2}{dt^2}g_p(t) + \omega^2_p(t) g_{p}(t), \label{eq.dimful.mode.eq} \\
 \omega^2_{p}(t) &=& |\vec{p}|^2 + V''(\phi_{\mathrm{cl}}(t)),
\end{eqnarray}
where $g_{p}(t)$ satisfies the Wronskian condition (alternatively the canonical commutation relation),
\begin{eqnarray}
 \dot{g}_p(t)g^{*}_p(t) - g_{p}(t)\dot{g}^{*}_p(t) = -i, \label{eq.Wronskian.dimful}
\end{eqnarray}
and coherent state $|\Omega\rangle$ is defined by the condition
\begin{eqnarray}
	a_{\vec{p}}|\Omega \rangle = 0.
\end{eqnarray}
One of most popular method to solve this equation is the Wentzel-Kramers-Brillouin (WKB) approximation \cite{BD}, namely, the mode function can be approximated by
\begin{eqnarray}
	g_p(t) = \frac{e^{-i\int^{t}W_p(t')dt'}}{\sqrt{2W_p(t)}},
\end{eqnarray}
where $W_p(t)$ satisfies the relation,
\begin{eqnarray}
 W^2_p(t) = \omega^2_p(t) 
 - \frac{1}{2}\left[ \frac{\ddot{W}_p(t)}{W_p(t)} - \frac{3}{2}\frac{\dot{W}^2_p(t)}{W^2_p(t)} \right].
\end{eqnarray}
If the dynamics of the system is quasi-static or adiabatic, $W_p(t)$ can be approximated by the series
\begin{eqnarray}
 W^2_p(t) \approx \omega^2_p(t) 
 \left[ 1 - \frac{\ddot{\omega}_p(t)}{2\omega^3_p(t)} + \frac{3\dot{\omega}^2_p(t)}{\omega^4_p(t)} \cdots \right],
\end{eqnarray}  
where the terms described by $\cdots$ are higher order terms. This method does not work for some cases, including (1) parametric amplification when $g_p(t)$ has an exponential growth and (2) spinodal instability for spontaneous symmetry breaking when $\omega^2_p(t)$ becomes negative \cite{Herring.2024}. 

The authors in Ref.~\cite{Herring.2024} discussed how parametric amplification appear as function of parameters in mode equation by the potential $V''(\phi)=m^2+3\lambda\phi^2$ with a mass $m$, a self coupling $\lambda$, and a simple classical field $\phi_{\mathrm cl}(t)=\phi(0)\cos(mt)$ to test their new formalism of the time-dependent-effective potential with the correction originating from the particle creation. The parametric amplification is described by the exponential growth in the Floquet's theory for Mathieu equation \cite{Gradshteyn, Abramowitz, math.dlmf}. In our case, we must extend this simple trigonometric function to more general periodic function,  the square of the Jacobi's elliptic functions, and discuss the stability based on so-called Hill's equation \cite{Hill.eq.book.1, Hill.eq.review.1}. We will discuss the parametric amplification later in the next section and will discuss the spinodal instability here.

It is best to convert our mode equation into dimensionless form. Introducing the dimensionless time variable, $\tau \equiv m_{\mathrm cl}t$, then we can rewrite the mode equation into the dimensionless mode function $\tilde{g}_p(\tau)$ with the dimensionless time $\tau$ with the dimensionless angular frequency $\tilde{\omega}_{p}(\tau) = \omega_p(\tau)/m_{\mathrm cl}(\phi_0)$:
\begin{eqnarray}
 0 =	\frac{d^2}{d\tau^2}\tilde{g}_{p}(\tau) + \tilde{\omega}^2_p(\tau) \tilde{g}_{p}(\tau), \label{eq.dimless.mode.eq}
\end{eqnarray}
where the dimensionless mode function $\tilde{g}_p$ is defined through the relation $\tilde{g}_p(\tau)=g_p(\tau)/\Lambda$ with a mass scale $\Lambda$.
Rewriting the Wronskian condition by the dimensionless quantities, we obtain
\begin{eqnarray} 
 \left[\frac{d}{d\tau}\tilde{g}_p(\tau)\right]\tilde{g}^{*}_p(\tau) - \tilde{g}_p(\tau)\left[\frac{d}{d\tau}\tilde{g}^{*}_p(\tau)\right] = -i\frac{\Lambda^2}{m_{\mathrm cl}}.
\end{eqnarray}
It is obvious that the choice, $\Lambda=\sqrt{m_{\mathrm cl}}$, enables us to obtain the same Wronskian condition for the dimensionless mode function $\tilde{g}_p(\tau)$. (See Ref.~\cite{KI.PLB.2024} for $m_{\mathrm cl}$.)

Using the potential contribution $V''(\phi_{\mathrm cl})$ in our time-dependent angular frequency $\omega^2_p(t)$:
\begin{eqnarray}
	V''(\phi_{\mathrm cl}) = -\mu^2 + 3\lambda \phi^2_{\mathrm cl}(t),
\end{eqnarray}
the dimensionless time-dependent angular frequency $\tilde{\omega}_p(\tau)$ reduces to
\begin{eqnarray}
	\tilde{\omega}^2_p(\tau) 
	&=& \frac{1}{m^2_{\mathrm cl}} 
	\left[	\vec{p}^2 + \lambda v^2 \left(-1 + 3\tilde{\phi}^2_{\mathrm cl}(\tau) \right)
	\right] \nn\\
	&\equiv& \tilde{p}^2 - \frac{4}{3}\alpha + 4\alpha \tilde{\phi}^2_{\mathrm cl}(\tau),    
\end{eqnarray}
with $\tilde{\phi}_{\mathrm cl}(t)=\phi_{\mathrm cl}(t)/v$ and we defined the dimensionless parameters:
\begin{eqnarray}
	\tilde{p}^2 &=& \frac{\vec{p}^2}{m^2_{\mathrm cl}}, \hspace{1cm}
	\alpha 
	\equiv \frac{3\lambda v^2}
	{4 m^2_{\mathrm cl}}.
\end{eqnarray} 
Note that we cannot regard $\alpha$ as a free parameter because $m_{\mathrm cl}$ is the function of $\tilde{\phi}_0$. This part is one of important difference between the our analysis of the mode equation and the analysis in Ref.~\cite{Herring.2024} for stability/instability. 

\begin{figure}[h]
 \includegraphics[scale=0.65]{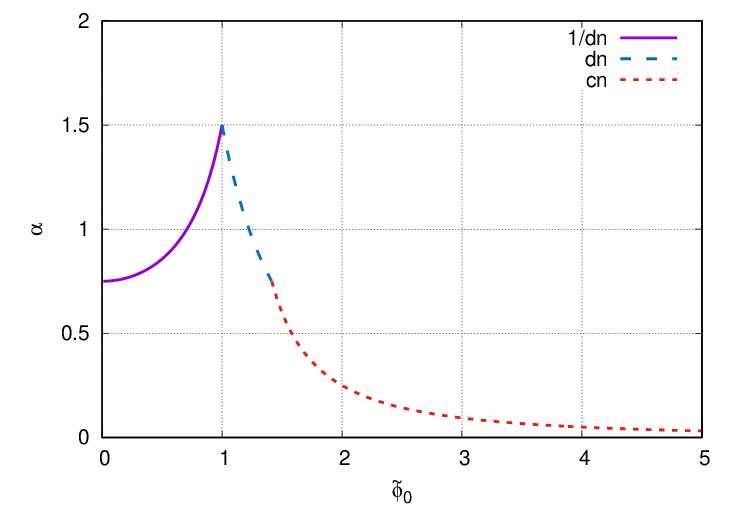}
  \caption{The parameter $\alpha$ as function of $\tilde{\phi}_0$. The solid (magenta in color), dashed (blue in color), and dotted (red in color) lines correspond to the parameter region of each mode, $1/dn$, $dn$ and $cn$, respectively. \label{fig.alpha}}
\end{figure}

We can rewrite $\alpha$ as a function $\tilde{\phi}_0$ by using the expression of $m_{\mathrm cl}$ obtained in \cite{KI.PLB.2024}:
\begin{equation}
	m^2_{\mathrm cl} =
	\begin{cases}
		\frac{\lambda v^2}{2}(2-\tilde{\phi}^2_0) & \text{$(0 < \tilde{\phi}_0 < 1)$},\\
		\frac{\lambda v^2}{2}\tilde{\phi}^2_0 &
		\text{$(1 < \tilde{\phi}_0 < \sqrt{2})$},\\
		2\lambda v^2 \left(\frac{\tilde{\phi}^2_0-1}{2}\right) & \text{$(\sqrt{2} < \tilde{\phi}_0)$},
	\end{cases}
\end{equation}
then $\alpha$ reduces to the simple function $\tilde{\phi}_0$:
\begin{equation}
	\alpha =
	\begin{cases}
		\frac{3}{2}\frac{1}{2-\tilde{\phi}^2_0} & \text{$(0 < \tilde{\phi}_0 < 1)$},\\
		\frac{3}{2}\frac{1}{\tilde{\phi}^2_0} & \text{$(1 < \tilde{\phi}_0 < \sqrt{2})$},\\
		\frac{3}{4}\frac{1}{\tilde{\phi}^2_0-1} & \text{$(\sqrt{2} < \tilde{\phi}_0)$}.
	\end{cases}
\end{equation}
The numerical plot of $\alpha$ is shown in Fig.~\ref{fig.alpha}. We can easily observe that the coefficient $\alpha$ is bounded between $0~(\tilde{\phi}_0 \to \infty)$ and $1.5~(\tilde{\phi}_0 \to 1)$. 
Then we can obtain $\tilde{\omega}^2_p(\tau)$ as:
\begin{eqnarray}
	\tilde{\omega}^2_p(\tau)
	= \tilde{p}^2 
	 + \left( - \frac{4}{3} + 4\tilde{\phi}^2_{\mathrm cl}(\tau)\right)\times
	\left\{
	\begin{array}{cc}
		\frac{3}{2}
		\frac{1}{2-\tilde{\phi}^2_0} & 
		(0 < \tilde{\phi}_0 < 1),\\
		\frac{3}{2}
		\frac{1}{\tilde{\phi}^2_0}
		& 
		(1 < \tilde{\phi}_0 < \sqrt{2}),\\
		\frac{3}{4}
		\frac{1}{\tilde{\phi}^2_0-1}
		& 
		(\sqrt{2} < \tilde{\phi}_0).
	\end{array}
	\right.
\end{eqnarray}
To discuss the stability/instability of the mode equation $\tilde{g}_p(\tau)$, we must analyze the nature of the time dependent angular frequency $\tilde{\omega}_p(\tau)$.

\begin{figure}[htb]
	\includegraphics[scale=0.65]{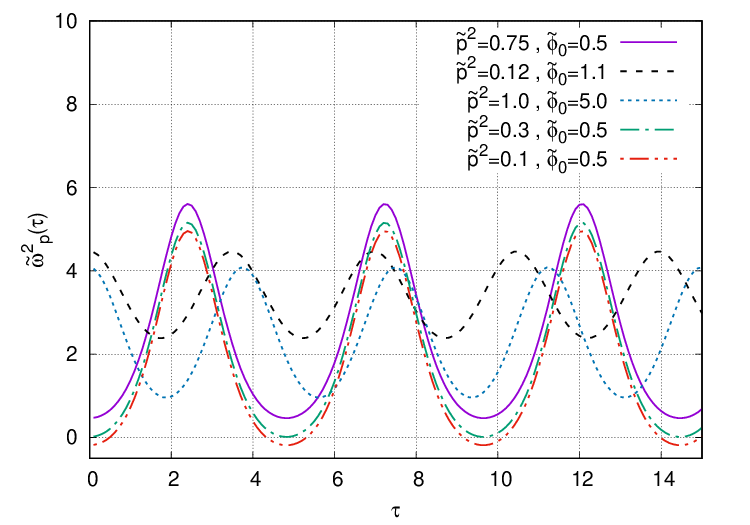}
	\caption{The behavior of $\tilde{\omega}^2_p$ as function of $\tau$ for various $\tilde{\phi}_0$ and $\tilde{p}^2$.}
	\label{fig.omega2.osci}
\end{figure}

First we check the numerical plot of $\tilde{\omega}^2_{p}(\tau)$ for given $\tilde{\phi}_0$ shown in Fig.~\ref{fig.omega2.osci}.
We can observe that the line (dot-solid-dot-dot-dot, red in color) ($\tilde{p}^2=0.1$, $\tilde{\phi}_0 = 0.5$) reaches negative region, while others oscillate in positive region. The parameter with $\tilde{\omega}^2_p(\tau)<0$ indicates a spinodal instability for the mode function. Which $\tilde{\phi}_0$ parameter gives this instability ? We can estimate the parameter region by obtaining the minimum value of $\tilde{\omega}^2_{p}$, denoted by $\tilde{\omega}^2_{p,\min}$. If this minimum value reaches negative region in a certain period, then the mode function shows instability. 

\begin{figure}[htb]
	\begin{tabular}{cc}
		\begin{minipage}[b]{0.55\linewidth}
			\centering
\includegraphics[scale=0.25]{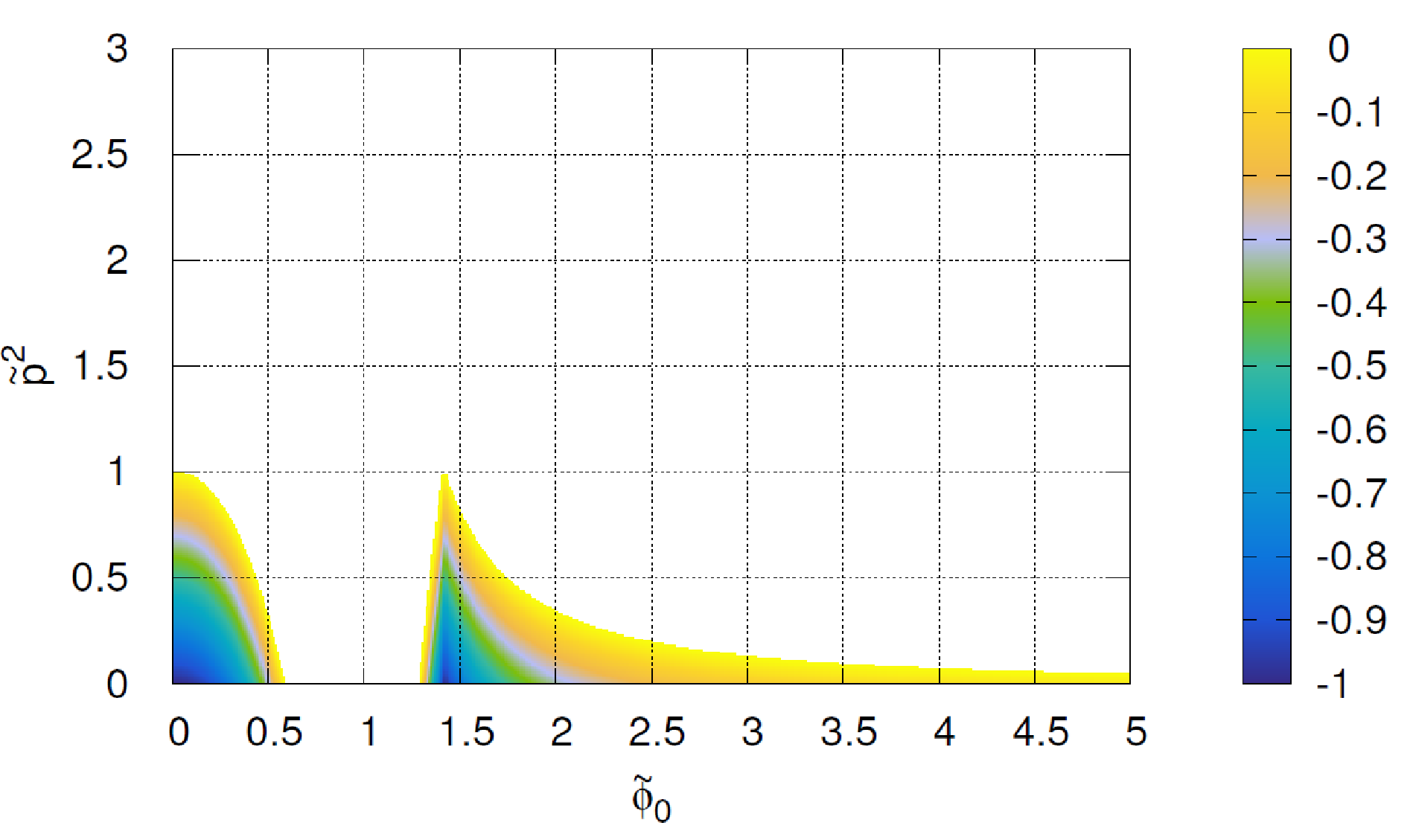}\\
			(a)  $\tilde{\omega}^2_{\min}<0$ in $(\tilde{\phi}_0, \tilde{p}^2)$
		\end{minipage}
		&
		\begin{minipage}[b]{0.55\linewidth}
			\centering
\includegraphics[scale=0.25]{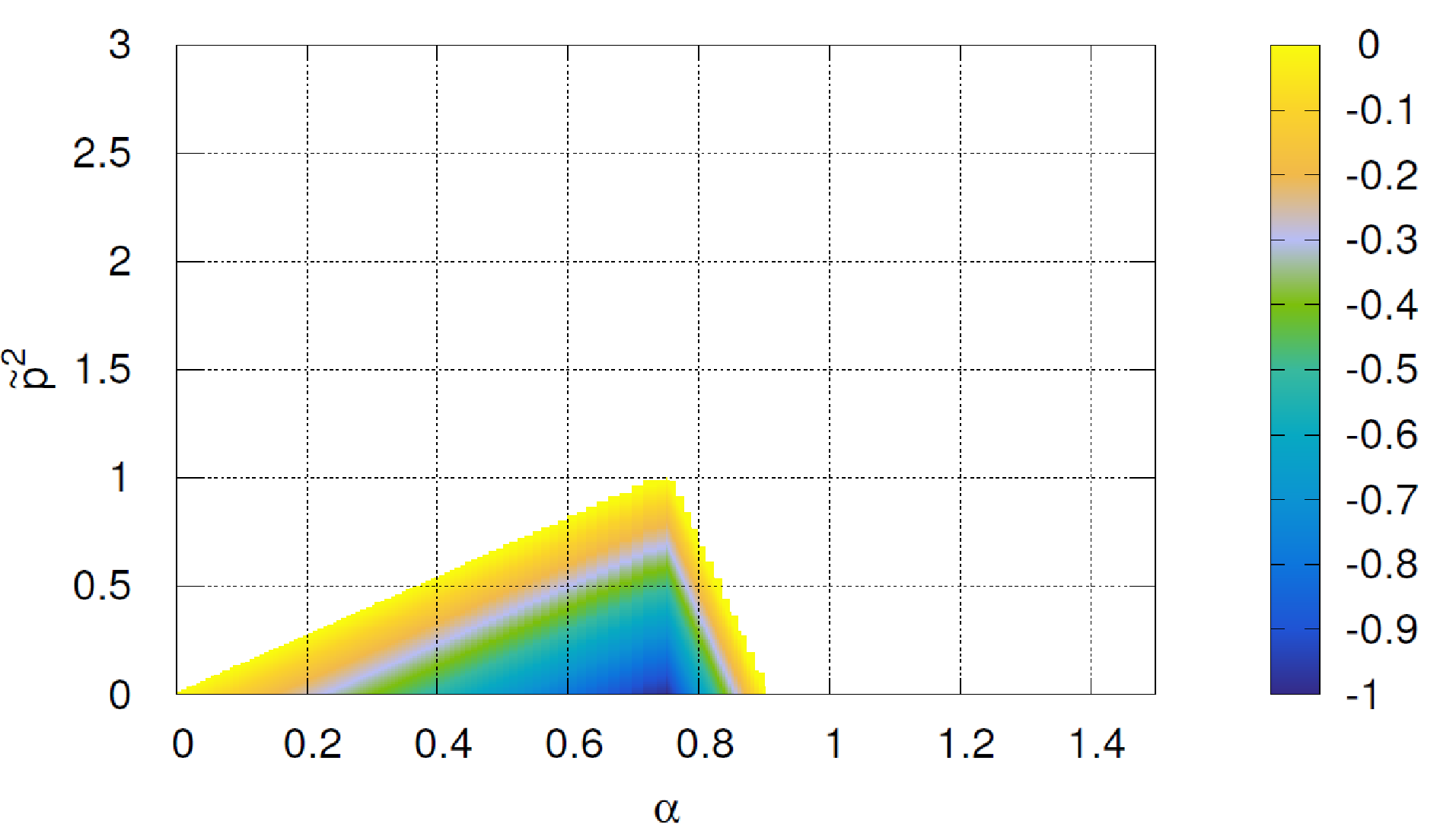}\\
			(b)  $\tilde{\omega}^2_{\min}<0$ in $(\alpha, \tilde{p}^2)$
		\end{minipage}
	\end{tabular}
	\caption{The region where $\tilde{\omega}^2_{\min}(\tau)<0$.} 
	\label{fig.omega2min.negative.region.phi0.alpha}
\end{figure}

Using the mathematical formula of $1/dn$, $dn$, $cn$ functions \cite{Abramowitz, Gradshteyn, math.dlmf}:
\begin{eqnarray}
	1 &\le& \frac{1}{dn(z,k)} \le \frac{1}{\sqrt{1-k^2}}, \nn\\
	\sqrt{1-k^2} &\le& dn(z,k) \le 1, \nn\\
	-1 &\le& cn(z,k) \le +1,
\end{eqnarray}
and the expression of $k_{1}$, $k_{2}$ and $k_{3}$ in Ref.~\cite{KI.PLB.2024}, we obtain
\begin{eqnarray}
	\tilde{\phi}_{\mathrm cl,\min} \le \tilde{\phi}_{\mathrm cl}(\tau) \le \tilde{\phi}_{\mathrm cl,\max},
\end{eqnarray}
with 
\begin{eqnarray}
\tilde{\phi}_{\mathrm cl, \min} 
 = 	\left\{
 \begin{array}{cc}
 \tilde{\phi}_0 & 
 	(0 < \tilde{\phi}_0 < 1),\\
    \sqrt{2-\tilde{\phi}^2_0}	& 
 	(1 < \tilde{\phi}_0 < \sqrt{2}),\\
 	-\tilde{\phi}_0	& 
 	(\sqrt{2} < \tilde{\phi}_0),
 \end{array}
 \right.
\end{eqnarray}
and 
\begin{eqnarray}
	\tilde{\phi}_{\mathrm cl, \max} 
	= 	\left\{
	\begin{array}{cc}
		\sqrt{2-\tilde{\phi}^2_0} & 
		(0 < \tilde{\phi}_0 < 1),\\
		\tilde{\phi}_0	& 
		(1 < \tilde{\phi}_0 < \sqrt{2}),\\
		+\tilde{\phi}_0	& 
		(\sqrt{2} < \tilde{\phi}_0).
	\end{array}
	\right.
\end{eqnarray}
Substituting these results into $\tilde{\omega}^2_{p}(\tau)$, we can obtain the minimum of this quantity for arbitrary time $\tau$:
\begin{eqnarray}
 \tilde{\omega}^2_{p,\min} 
 &=&
 	\left\{
 \begin{array}{cc}
 	\tilde{p}^2 + \frac{-2+6\tilde{\phi}^2_0}{2-\tilde{\phi}^2_0} & 
 	(0 < \tilde{\phi}_0 < 1),\\
 	\tilde{p}^2 + \frac{10-6\tilde{\phi}^2_0}{\tilde{\phi}^2_0} & 
 	(1 < \tilde{\phi}_0 < \sqrt{2}),\\
 	\tilde{p}^2 - \frac{1}{\tilde{\phi}^2_0-1}	& 
 	(\sqrt{2} < \tilde{\phi}_0).
 \end{array}
 \right.
\end{eqnarray}
As is shown in Fig.~\ref{fig.omega2min.negative.region.phi0.alpha}, we can plot the region where satisfies the condition $\tilde{\omega}^2_{p,\min}<0$ in $(\tilde{\phi}_0,\tilde{p}^2)$-plane and $(\alpha,\tilde{p}^2)$-plane. The colored region in (a) and (b) in Fig.~\ref{fig.omega2min.negative.region.phi0.alpha} correspond to the candidate of the unstable region.

\section{Instability based on discriminant of Hill's equation \label{sec3}}
The analysis based on $\tilde{\omega}^2_p$ is not full description of the instability, because it does not take into account the parametric amplification. The systematic way is to use the mathematical knowledge of the differential equation in Eq.~(\ref{eq.dimless.mode.eq}), namely, the property of Hill's equation.

\subsection{Hill's equation }
Our mode equation in Eq.~(\ref{eq.dimless.mode.eq}) reduces to the so-called Hill's equation \cite{Hill.eq.book.1,Hill.eq.review.1}. The Hill's equation is described by 
\begin{eqnarray}
 \frac{d^2}{dt^2}y(t) + a(t)y(t) = 0,
\end{eqnarray}
where the function $a(t)$ is a periodic function with a period $T$; $a(t+T)=a(t)$. The periodicity in $a(t)$ seems to imply an existence of a periodic solution $y(t+T)=y(t)$ at first glance. However, according to the general theory of Hill's equation, alternatively, Floquet's theory~\cite{Hill.eq.book.1, Hill.eq.review.1}, we can find out the solution which satisfies the property,
\begin{eqnarray}
	y(t+T)=\rho y(t), \label{eq.def.rho}
\end{eqnarray}
where $\rho$ is a constant (complex in general). The stability of solution in the Hill's equation equals to the boundedness of the solutions under a given initial condition. It is obvious that the solution $y(t)$ satisfies $y(t+2T)=\rho^2 y(T)$ and $y(t+3T)=\rho^3y(t), \cdots$, thus after $n$-times we obtain the result,
\begin{eqnarray}
	y(t+nT)=\rho^n y(t).
\end{eqnarray}
Taking a large $n$ value and setting $t=0$, the above equation indicates that we can determine the behavior of the solution at $t=\infty$, i.e., $y(t=\infty)$, in terms of the magnitude of $|\rho|$ and the initial value of $y(0)$; namely, if $|\rho|>1$ then the solution is unbounded (unstable), if $|\rho|<1$ then the solution is bounded (stable), and if $|\rho|=1$ then the solution reduces to the purely periodic or anti-periodic function (bounded). 

Moreover, we can derive the characteristic equation:
\begin{eqnarray}
	\rho^2 - \left(u_1(T)+u'_2(T)\right)\rho + 1 = 0, \label{eq.characteristic}
\end{eqnarray}
where the $u_1(t)$ and $u_2(t)$ are called normalized solutions which satisfy the initial conditions:
\begin{eqnarray}
	 u_1(0) = 1, \hspace{1cm} u_2(0) = 0,\hspace{1cm}
	u'_1(0) = 0, \hspace{1cm} u'_2(0) = 1.
\end{eqnarray}
Let $\rho_1$ and $\rho_2$ be two solutions of the above characteristic equation for $\rho$ in Eq.~(\ref{eq.characteristic}), then obviously, these $\rho_1$ and $\rho_2$ satisfy the two conditions:
\begin{eqnarray}
	\rho_1+\rho_2 = u_1(T)+u'_2(T), \hspace{1cm} \rho_1 \rho_2 = 1.
\end{eqnarray}
One can easily obtain the discriminant of the characteristic equation in Eq.~(\ref{eq.characteristic}), denoted by $D$, is given in:
\begin{eqnarray}
 D = (u_1(T) + u'_2(T))^2-4 \equiv d^2-4, \label{eq.discriminant}
\end{eqnarray} 
where we defined $d=|u_1(T) + u'_2(T)|$ and we call it the discriminant of the Hill's equation.

According to the Floquet's theory, the two linearly independent solutions of the Hill's equation denoted by $y_1$ and $y_2$ can be expressed by
\begin{eqnarray}
 y_1(t) = e^{+i\nu t}\Pi_1(t), \hspace{1cm}
 y_2(t) = e^{-i\nu t}\Pi_2(t),
\end{eqnarray}
where the index $\nu$ is called Floquet'ts exponent, $\Pi_1$ and $\Pi_2$ are $nT$-periodic functions with $n\neq 1, 2$, and the relations $\rho_1=e^{-i\nu T}$ and $\rho_2=e^{i\nu T}$ hold.
Furthermore, it is known that: (1) if $d>2$, then $\rho_1$ and $\rho_2$ reduce to real numbers and thus the solution is unstable (unbounded), (2) if $d<2$, then $\rho_1$ and $\rho_2$ reduce to complex numbers and thus the solution is stable (bounded), and (3) if $d=2$, then $\rho=\pm 1$ and thus the solution reduces to $T$-periodic or $T$-anti-periodic stable solution.  

\subsection{Stability of our mode equation}
\begin{figure}[htb]
	\includegraphics[scale=0.65]{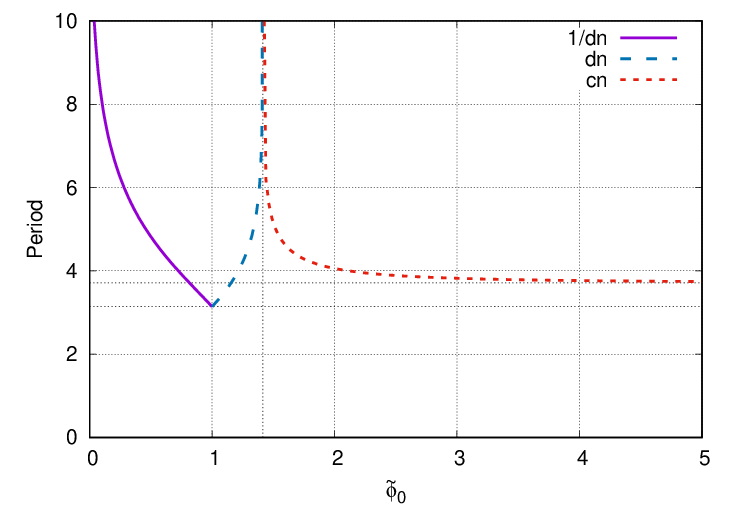}
	 \caption{The period $T$ of the time dependent angular frequency $\tilde{\omega}^2(\tau)$ as function of $\tilde{\phi}_0$.  The solid (magenta in color), dashed (blue in color), and dotted (red in color) lines corresponds to the parameter region for $1/dn$, $dn$ and $cn$ modes, respectively. The period $T$ approaches to the constant, $2K(1/\sqrt{2})\approx 3.71$ for $\tilde{\phi}_0 \to \infty$; while $T$ approaches to $\pi\approx 3.14$ for $\tilde{\phi}_0 \to 1$. }
	\label{fig.period.T}
\end{figure}

To use the stability condition of Hill's equation, we must determine the period of $\tilde{\omega}^2(\tau)$.
We can easily find out the period of the time-dependent angular frequency $\tilde{\omega}^2(\tau)$ by using the mathematical identities \cite{Gradshteyn, Abramowitz, math.dlmf};  
\begin{eqnarray}
dn^2(z+2K(k),k)=dn^2(z,k), \hspace{1cm} cn^2(z+2K(k),k)=cn^2(z,k),
\end{eqnarray}
where $K(k)$ is the complete elliptic integral of the first kind. Thus the period $T$ in $\tilde{\omega}^2_p(\tau)$ in the mode equation is given in:
\begin{eqnarray}
 T = 
  \left\{
 \begin{matrix}
 	2K(k_1) \hspace{1cm}(0<\tilde{\phi}_0 \le 1),\\
 	2K(k_2) \hspace{1cm}(v < \tilde{\phi}_0 < \sqrt{2}),\\
 	2K(k_3) \hspace{1cm}(\sqrt{2} < \tilde{\phi}_0),
 \end{matrix}
 \right.
\end{eqnarray}
where $k_{1}$, $k_{2}$ and $k_{3}$ are obtained in Ref.~\cite{KI.PLB.2024}. The numerical plot of the period $T$ as the function of $\tilde{\phi}_0$ is shown in Fig.~\ref{fig.period.T}.
Note that the period $T$ diverges for $\tilde{\phi}_0\to 0$ and $\tilde{\phi}_0 \to \sqrt{2}$. 
This is because $V(\tilde{\phi}=0)=V(\tilde{\phi}_0=\sqrt{2})$, thus the transition of the field value $\tilde{\phi}_{\mathrm cl}$ takes infinite time to change from one to another. Therefore we exclude these parameters $\tilde{\phi}_0=0$ and $\tilde{\phi}_0 =\sqrt{2}$. 

\begin{figure}[htb]
	\begin{tabular}{cc}
		\begin{minipage}[b]{0.55\linewidth}
			\centering
	\includegraphics[scale=0.26]{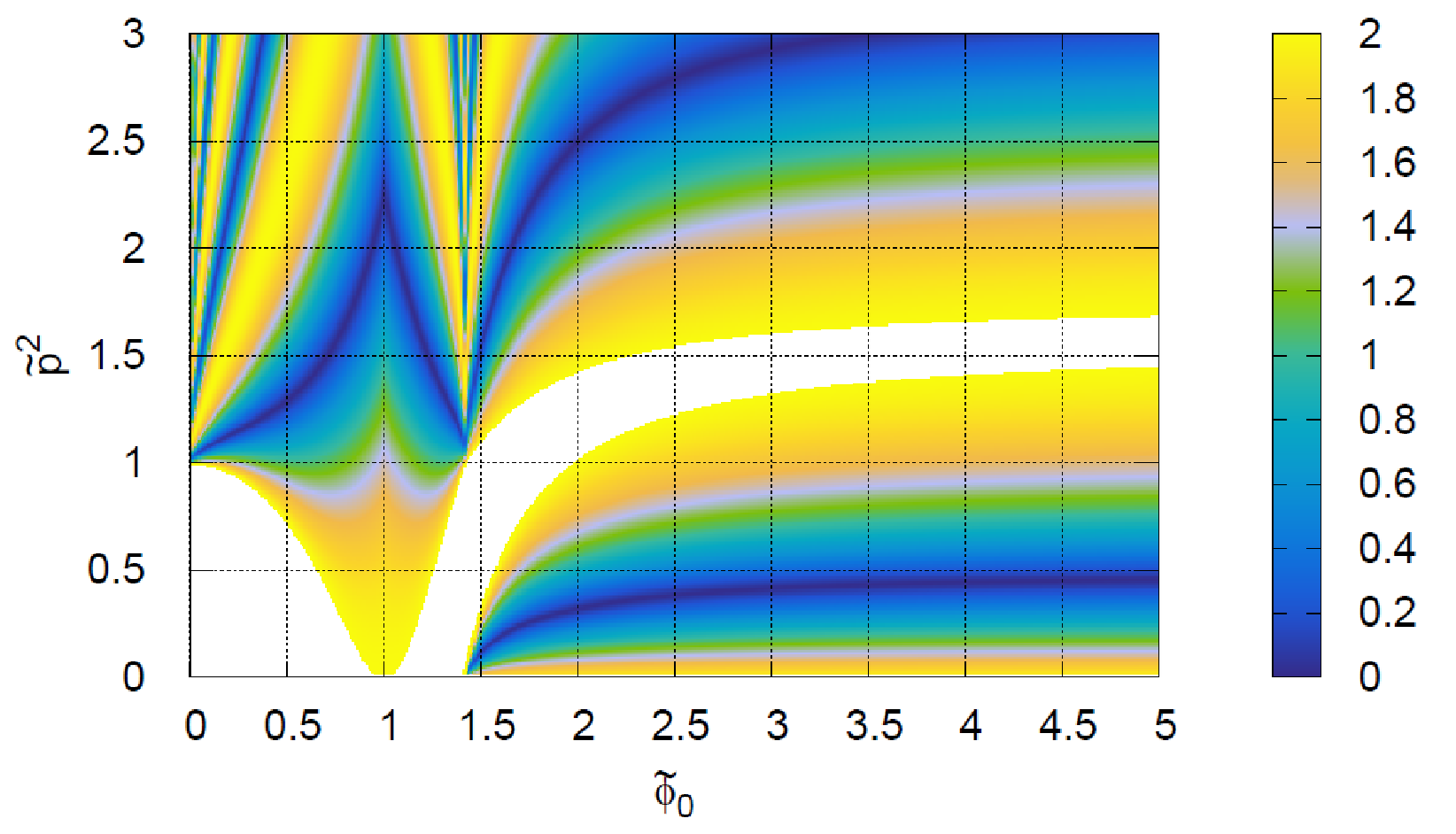}\\
			(a) $(\tilde{\phi}_0,\tilde{p}^2)$ plane
		\end{minipage}
		&
		\begin{minipage}[b]{0.55\linewidth}
			\centering
	\includegraphics[scale=0.22]{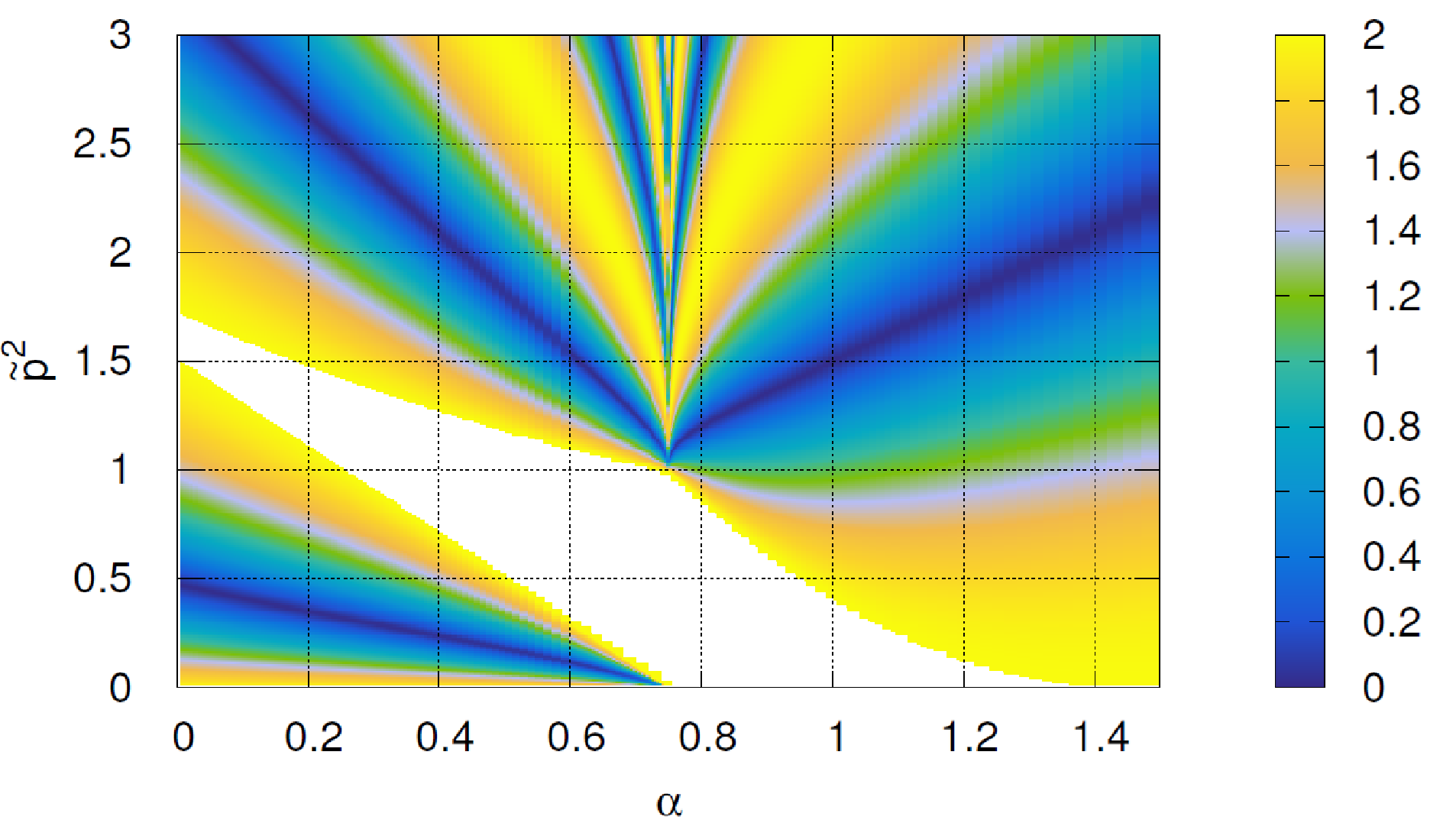}\\
			(b) $(\alpha,\tilde{p}^2)$ plane
		\end{minipage}
	\end{tabular}
	\caption{The region where the discriminant of the Hill's equation satisfy $d<2$ as function of parameters in the mode equation. The numerical value of $d$ is projected on the two planes. The uncolored region (in color) corresponds to the region with $d>2$.} 
	\label{fig.discriminant.larger.than.2.p2.phi0.alpha.space}
\end{figure}

Solving the mode equation numerically and calculating $u_1(T)$ and $u'_2(T)$, we can obtain the value of $d$ in $(\tilde{\phi}_0,\tilde{p}^2)$-plane and $(\alpha,\tilde{p}^2)$-plane. The results of the stable regions where $d=|u_1(T) + u'_2(T)|<2$ are shown in Fig.~\ref{fig.discriminant.larger.than.2.p2.phi0.alpha.space}.
We can easily observe that the result of unstable region (uncolored region in color) in Fig.~\ref{fig.discriminant.larger.than.2.p2.phi0.alpha.space} not only cover the unstable region near the origin in $(\tilde{\phi}_0,\tilde{p}^2)$ plane in Fig.~\ref{fig.omega2min.negative.region.phi0.alpha} which is suggested by the analysis for $\tilde{\omega}^2_{\min}<0$, but also reveals other unstable region (band). These band regions both include the spinodal instability and the parametric amplification. Because the result gives the fundamental nature to discuss the instability of the quantum mode, this is one of main findings in our article.

\subsection{Solutions of mode equation for stable/unstable parameters}
We plot the dimensionless mode function $\tilde{g}_{p}(\tau)$ for stable/unstable parameters based on the results in the previous subsection and in the result of Fig.~\ref{fig.discriminant.larger.than.2.p2.phi0.alpha.space}.
Note that the dimensionless mode function $\tilde{g}_p(\tau)$ which satisfies the condition:
\begin{eqnarray}
 \tilde{g}_{p}(\tau) &=& au_1(\tau) + bu_2(\tau),
\end{eqnarray}
where the above coefficients stand for the initial conditions, $\tilde{g}_p(0)=a$ and $\frac{d\tilde{g}_p(\tau)}{d\tau}\large|_{\tau=0}=b$. These complex coefficients $a$ and $b$ satisfy the Wronskian condition, $ ba^{*}-ab^{*}=-i$.
We follow the choice of the initial condition in Ref.~\cite{Herring.2024}: 
\begin{eqnarray}
 a=\frac{1}{\sqrt{2\tilde{\omega}_p(0)}}, \hspace{1cm}
 b=\frac{-i\tilde{\omega}_p(0)}{\sqrt{2\tilde{\omega}_p(0)}}.
\end{eqnarray}
We take the parameters near the instability band (uncolored region in Fig.~\ref{fig.discriminant.larger.than.2.p2.phi0.alpha.space}) to check the behavior of the mode functions with stability/instability. 

First, we see the parameter set which gives the stable solution:
\begin{eqnarray}
 (a)~\mbox{stable~$1/dn$}:&~&(\tilde{p}^2=0.75,\tilde{\phi}_0=0.5), \nn\\
 (b)~\mbox{stable~$dn$}:&~&(\tilde{p}^2=0.12,\tilde{\phi}_0=1.1), \nn\\
 (c)~\mbox{stable~$cn$}:&~&( \tilde{p}^2=1.0, \tilde{\phi}_0=5.0).\nn
\end{eqnarray}
\begin{figure}[htb]
	\begin{tabular}{ccc}
		\begin{minipage}[b]{0.35\linewidth}
			\centering
\includegraphics[scale=0.45]{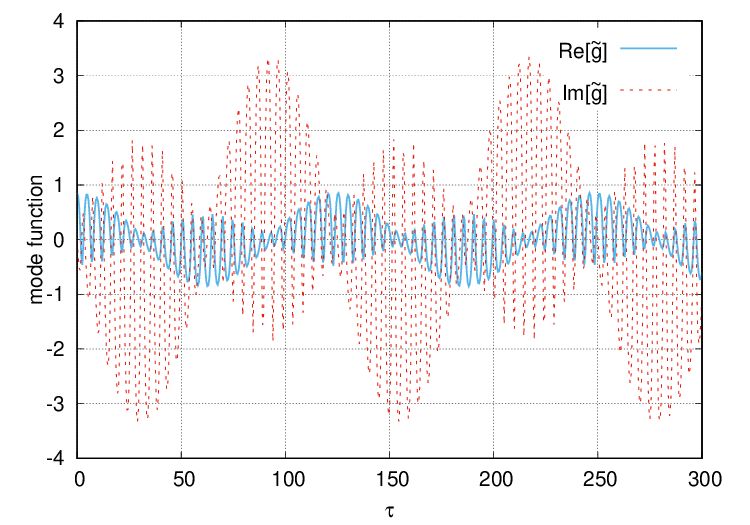}\\
			(a) $1/dn$ region
		\end{minipage}
		&
		\begin{minipage}[b]{0.35\linewidth}
			\centering
\includegraphics[scale=0.45]{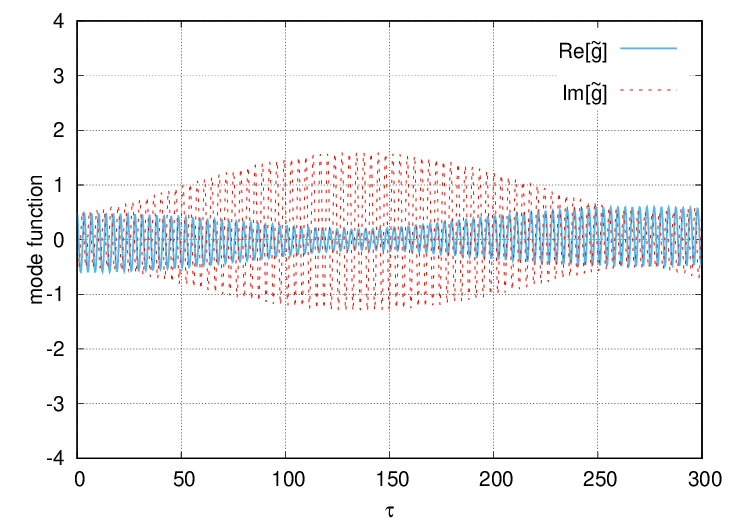}\\
(b) $dn$ region
		\end{minipage}
	    &
\begin{minipage}[b]{0.35\linewidth}
	\centering
\includegraphics[scale=0.45]{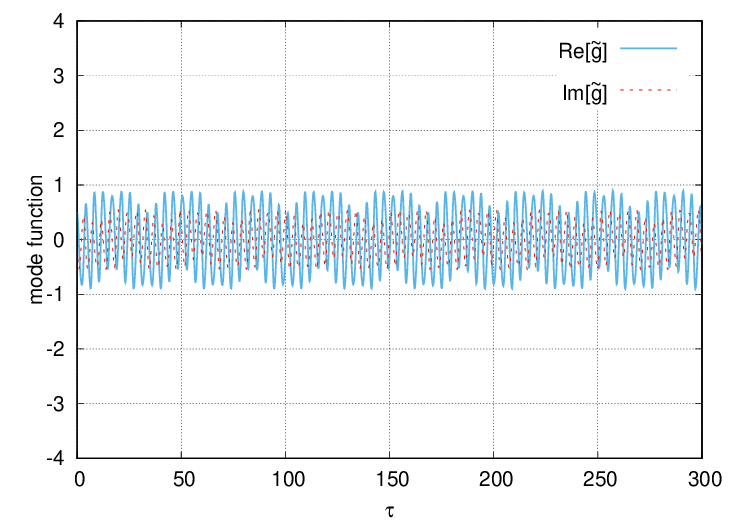}\\
(c) $cn$ region
\end{minipage}
	\end{tabular}
	\caption{The stable solutions for the three modes.} 
	\label{fig.mode.func.g.invdn.dn.cn.stable}
\end{figure}
The numerical plots of $\tilde{g}_p(\tau)$ with these three parameters are shown in Fig.~\ref{fig.mode.func.g.invdn.dn.cn.stable}. We explicitly checked these solutions satisfy the Wronskian condition for consistency, as expected. These graphs show that:  the mode function has two different oscillations, namely, one is very short and another is relatively longer (see (a) of Fig.~\ref{fig.mode.func.g.invdn.dn.cn.stable}) if the stable parameter is close to the instability band, while the mode function has almost one period (see (c) of Fig.~\ref{fig.mode.func.g.invdn.dn.cn.stable}) if the stable parameter deviates from the instability band. Moreover, the closer stable parameter to the band tend to give a larger amplitude of the solutions. These behaviors are consistent with the general theory of the Hill's equation \cite{Hill.eq.book.1, Hill.eq.review.1}. 

\begin{figure}[htb]
	\begin{tabular}{ccc}
		\begin{minipage}[b]{0.55\linewidth}
			\centering
\includegraphics[scale=0.6]{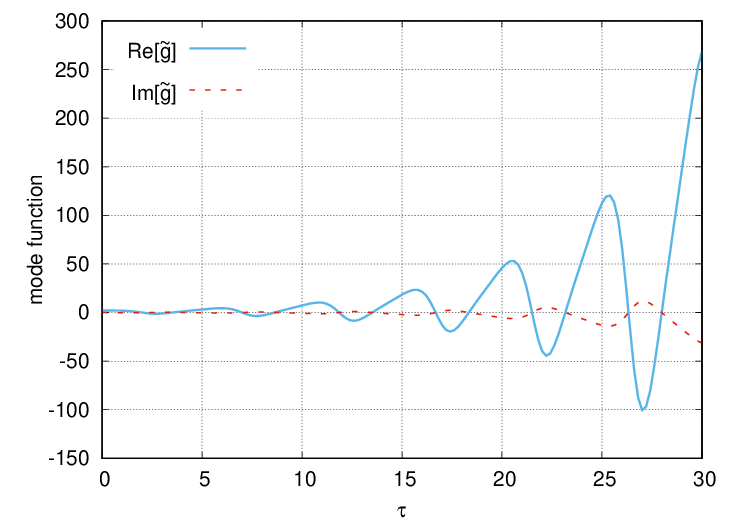}\\
(a) parametric amplification
		\end{minipage}
		&
		\begin{minipage}[b]{0.55\linewidth}
			\centering
\includegraphics[scale=0.6]{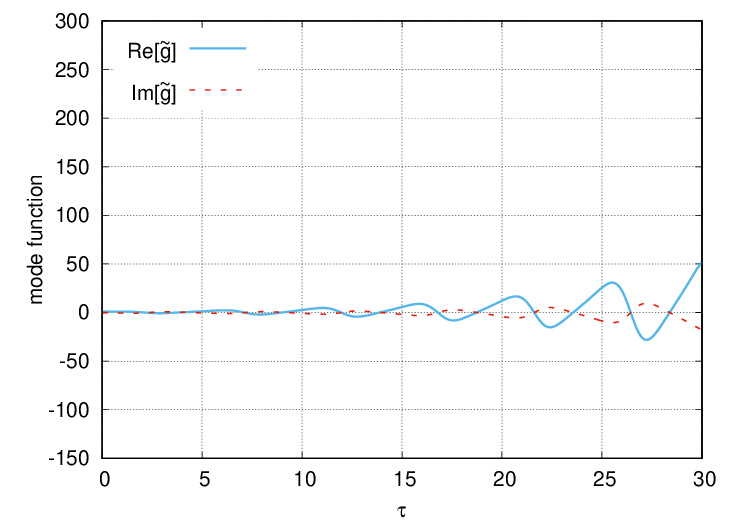}\\
(b) spinodal instability
		\end{minipage}
	\end{tabular}
	\caption{Two kinds of unstable solutions.} 
	\label{fig.mode.func.g.unstable}
\end{figure}

Next, we see the parameters which give the unstable solutions:
\begin{eqnarray}
	(a)~\mbox{parametric amplification}:&~&(\tilde{p}^2=0.3,\tilde{\phi}_0=0.5, d=2.7, T=4.8, \nu=0.17i), \nn\\
	(b)~\mbox{spinodal instability}:&~&(\tilde{p}^2=0.1,\tilde{\phi}_0=0.5, d=2.4, T=4.8, \nu=0.13i), \nn
\end{eqnarray}
where we can check these two points are located in the instability band in Fig.~\ref{fig.discriminant.larger.than.2.p2.phi0.alpha.space}.
Note that unstable solutions have two types of the instability; namely, 1) the first one is called parametric amplification, 2) the second one is called spinodal instability. The numerical plots of the solution are show in Fig.~\ref{fig.mode.func.g.unstable}. 

These results show that the mode functions show exponential growths with oscillations when we choose the unstable parameters. 
In particular, the particle number behaves like $\simeq e^{\pm 2i\nu\tau}$ with corresponding Floquet's exponent, because the mode function behaves like $\tilde{g}\simeq e^{\pm i\nu\tau}$ in the unstable parameter region, as we discuss later.

\begin{figure}[htb]
	\begin{tabular}{ccc}
		\begin{minipage}[b]{0.52\linewidth}
			\centering
\includegraphics[scale=0.25]{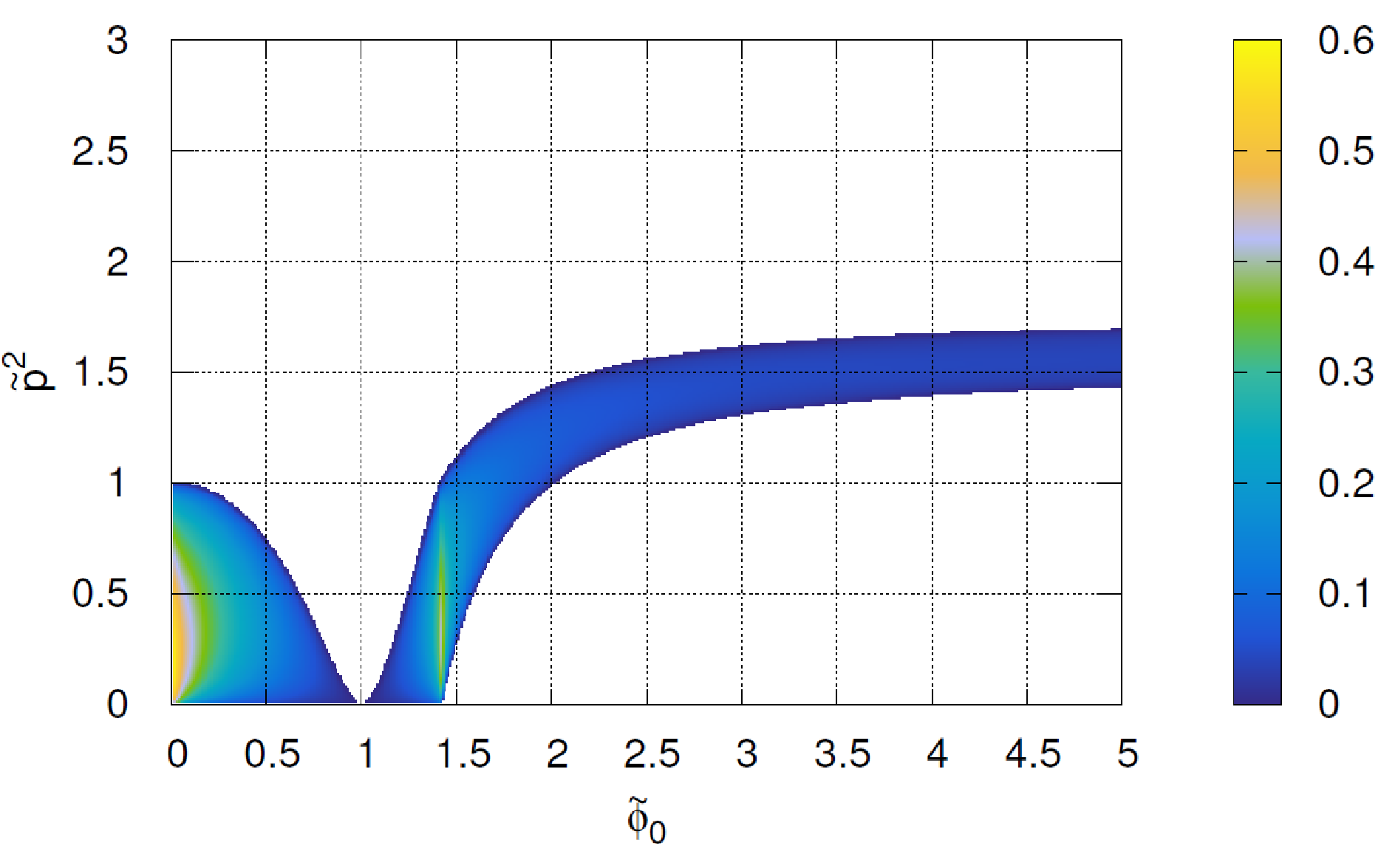}\\
		(a) $(\tilde{\phi}_0,\tilde{p}^2)$ plane
		\end{minipage}
		&
		\begin{minipage}[b]{0.52\linewidth}
			\centering
\includegraphics[scale=0.25]{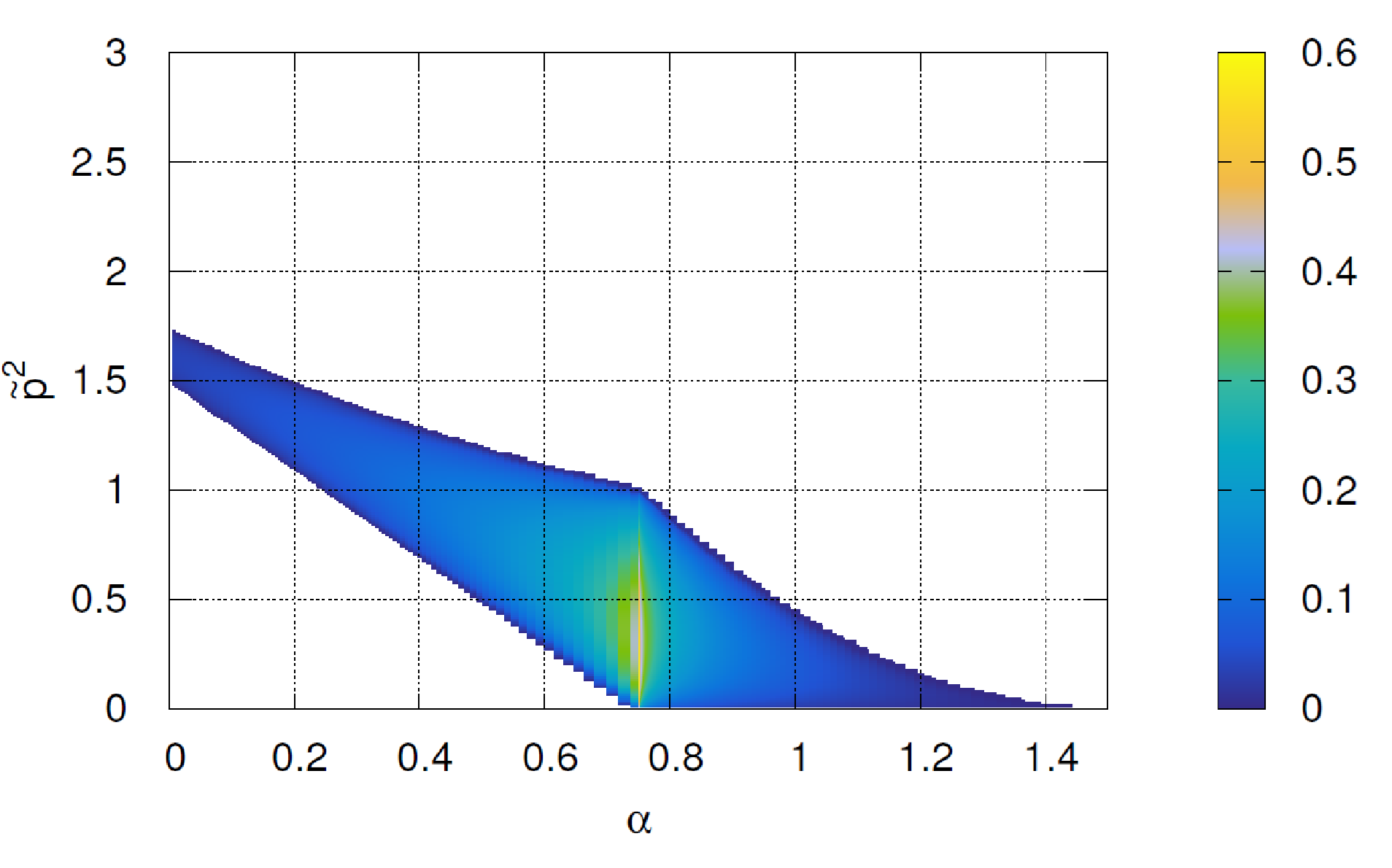}\\
			(b) $(\alpha,\tilde{p}^2)$ plane
		\end{minipage}
	\end{tabular}
	\caption{The non-zero imaginary part of the Floquet's exponent for the unstable parameter region.} 
	\label{fig.Floquet.exp.p2.phi0.alpha.space}
\end{figure}

The non-zero imaginary parts of the Floquet's exponent as the function of $(\tilde{\phi}_0,\tilde{p}^2)$ and $(\alpha,\tilde{p}^2)$ are shown in Fig.~\ref{fig.Floquet.exp.p2.phi0.alpha.space}.
Note that the two regions $\tilde{\phi}_0\approx 0$ and $\tilde{\phi}_0\approx \sqrt{2}$ in (a) and the region $\alpha\approx 0.75$ in (b) of Fig.~\ref{fig.Floquet.exp.p2.phi0.alpha.space} show the significant change of the behavior. This is because these parameter regions correspond to the edge of the parameters where the classical solutions loose the periodicity.

\subsection{Particle number for adiabatic particles}
Finally, we discuss the particle creation process through the classical field to adiabatic particles. First, adopting the zero-th order-adiabatic-mode as basis $\tilde{f}_p(t)$ in Ref.~\cite{Herring.2024}, we express the mode function $g_p(t)$ in terms of $\tilde{f}_p(t)$,
\begin{eqnarray}
g_p(t) 
 &=& \tilde{A}_p(t)\tilde{f}_p(t) + \tilde{B}_p(t)\tilde{f}^{*}_p(t),
\end{eqnarray} 
where we introduced the Bogoliubov coefficients $\tilde{A}_p(t)$ and $\tilde{B}_p(t)$ which satisfy the Wronskian condition,
\begin{eqnarray}
 |\tilde{A}_p(t)|^2 - |\tilde{B}_p(t)|^2 = 1.
\end{eqnarray} 
Next, we introduce the dimensionless zero-th-order-adiabatic-mode  $\tilde{F}_p(t)$,
\begin{eqnarray}
 \tilde{f}_p(t) 
 &=& \frac{e^{-i\int^{t}\omega_p(t')dt'}}{\sqrt{2\omega_p(\tau)}}
 = \frac{1}{\sqrt{m_{\mathrm cl}}} \frac{e^{-i\int^{\tau}\tilde{\omega}_p(\tau')d\tau'}}{\sqrt{2\tilde{\omega}_p(\tau)}} 
 \equiv \frac{\tilde{F}_{p}(\tau)}{\sqrt{m_{\mathrm cl}}},
\end{eqnarray}
where we rewrite dimensionful quantity $\tilde{f}_p(t)$ by the dimensionless quantity $\tilde{F}_p(\tau)$. Thus we obtain the relation between $\tilde{g}_p(\tau)$ and $\tilde{F}_p(\tau)$ as:
\begin{eqnarray}
	\tilde{g}_p(t) 
	&=& \tilde{A}_p(t)\tilde{F}_p(t) + \tilde{B}_p(t)\tilde{F}^{*}_p(t), \\
	\dot{\tilde{g}}_p(t) &=& -i\tilde{\omega}_p(t)\left[ \tilde{A}_p(t)\tilde{F}_p(t) - \tilde{B}_p(t)\tilde{F}^{*}_p(t)\right]. 
\end{eqnarray}
These parameters $\tilde{A}$ and $\tilde{B}$ mix the annihilation operator $a_{\vec{p}}$ with creation operators $a^{\dagger}_{\vec{p}}$ in the quantum fluctuation of the scalar field:
\begin{eqnarray}
	a_{\vec{p}}g_{p}(t) + a^{\dagger}_{-\vec{p}}g^{*}(t) 
	= c_{\vec{p}}(t)\tilde{f}_p(t) + c^{\dagger}_{-p}(t)\tilde{f}^{*}_p(t),
\end{eqnarray} 
and we define the new basis of time-dependent-operators $c_{\vec{p}}(t)$ and $c^{\dagger}_{\vec{p}}(t)$ as:
\begin{eqnarray}
 c_{\vec{p}}(t) &=& \tilde{A}_{p}(t)a_{\vec{p}}(t) + \tilde{B}^{*}_p(t)a^{\dagger}_{-\vec{p}}, \\
 c^{\dagger}_{-\vec{p}}(t) &=& \tilde{A}^{*}_{p}(t)a^{\dagger}_{-\vec{p}}(t) + \tilde{B}_p(t)a_{\vec{p}}, 
\end{eqnarray}
where $c$ and $c^{\dagger}$ satisfy the same commutation relations for $a$ and $a^{\dagger}$ and define the instantaneous adiabatic vacuum state $|0_a(t)\rangle$ at time $t$: $c_{\vec{p}}(t)|0_a(t)\rangle=0$.  

For $\tilde{\omega}^2_p(\tau)>0$ case including parametric amplification,
the number of the adiabatic particles, $\tilde{N}_{p}(t)$, created by the operator $c^{\dagger}$ at a given time for the coherent state $|\Omega\rangle$ can be obtained in terms of the coefficient $\tilde{B}_p(t)$
\begin{eqnarray}
 \tilde{N}_p(t) = \langle \Omega| c^{\dagger}_{\vec{p}}~ c_{\vec{p}}|\Omega\rangle = |\tilde{B}_p(t)|^2,
\end{eqnarray}
where the instantaneous adiabatic vacuum $|0_a(t)\rangle$ for the time-dependent operator $c_{\vec{p}}(t)$ and the coherent state $|\Phi\rangle$ for the time-independent operator $a_{\vec{p}}$ is related to each other through two-mode-squeezed state \cite{Herring.2024}. Alternatively, this can be expressed by $\tilde{g}_p(\tau)$ as 
\begin{eqnarray}
	\tilde{N}_p(t)
&=&\frac{1}{2\tilde{\omega}_p(\tau)}\left[ \Big|\frac{\tilde{g}_p(\tau)}{d\tau}\Big|^2 + \tilde{\omega}^2_p(\tau)|\tilde{g}_p(\tau)|^2\right] - \frac{1}{2},
\end{eqnarray}
where we substituted the relation $g_p(t)=\tilde{g}_p(\tau)/\sqrt{m_{\mathrm cl}}$ into the above definition. 

On the other hand, for $\tilde{\omega}^2_p(\tau)<0$ case, we need a modification of the definition. In our case, the time-dependent-angular-frequency shows such property for some spinodal instability. In this case, if the $\omega^2_p(\tau)$ becomes a negative value, we define the new time-dependent-dimensionless-angular-frequency, $\bar{\omega}_p(\tau)$:
\begin{eqnarray}
 \bar{\omega}_p(\tau) \equiv \sqrt{\tilde{p}^2 + |V''(\phi_{\mathrm cl}(\tau))|}, 	 
\end{eqnarray}
for the spinodal parameter and introduce the redefined expansion for $\tilde{g}_p(t)$ by
\begin{eqnarray}
 \tilde{g}_p(t) &=& \bar{A}_p(t)\bar{F}_p(t) + \bar{B}_p(t)\bar{F}^{*}_p(t), \\
 \dot{\tilde{g}}_p(t) &=& -i\bar{\omega}_p(t)\left[ \bar{A}_p(t)\bar{F}_p(t) - \bar{B}_p(t)\bar{F}^{*}_p(t)\right], 
\end{eqnarray}
with 
\begin{eqnarray}
 \bar{f}_p(t)=\frac{e^{-i\int^{t}\bar{\omega}_p(t')dt'}}{\sqrt{2\bar{\omega}_p(t)}} \equiv \frac{\bar{F}_p(\tau)}{\sqrt{m_{\mathrm cl}}}.
\end{eqnarray}
Then we can define the particle number $\bar{N}_p(t)$ for spinodal parameter region as the analogy of the parametric amplification case:
\begin{eqnarray}
 \bar{N}_p(\tau) = |\bar{B}_p(\tau)|^2 = \frac{1}{2\bar{\omega}_p(\tau)}
 \left[ \Big|\frac{d\tilde{g}_p(\tau)}{d\tau}\Big|^2  + \bar{\omega}^2_p(\tau)|\tilde{g}_p(\tau)|^2\right] - \frac{1}{2}.
\end{eqnarray}
Basically other definition of the particle/anti-particle is possible \cite{BD,Mukhanov.Winitzki.2012}. 
However, the advantages of these definition are discussed in \cite{Herring.2024}, namely, if there is an asymptotic stationary state, then the angular frequency $\omega_p(\infty)$ and annihilation/creation operators,  $c_{\vec{p}}(\infty)$ and $c^{\dagger}_{\vec{p}}(\infty)$ become constant in time. Also see Refs.~\cite{Parker.1968.to.2012,Ford.1987,Parker.Toms.2009,Habib.Molina-Paris.Mottola.1999,Dabrowwski.Dunne.2014} for the time-dependent-operators by expanding the basis of the zeroth-order adiabatic modes. 

\begin{figure}[htb]
			\centering
	\begin{tabular}{cc}
	\begin{minipage}[b]{0.55\linewidth}
		\centering
\includegraphics[scale=0.6]{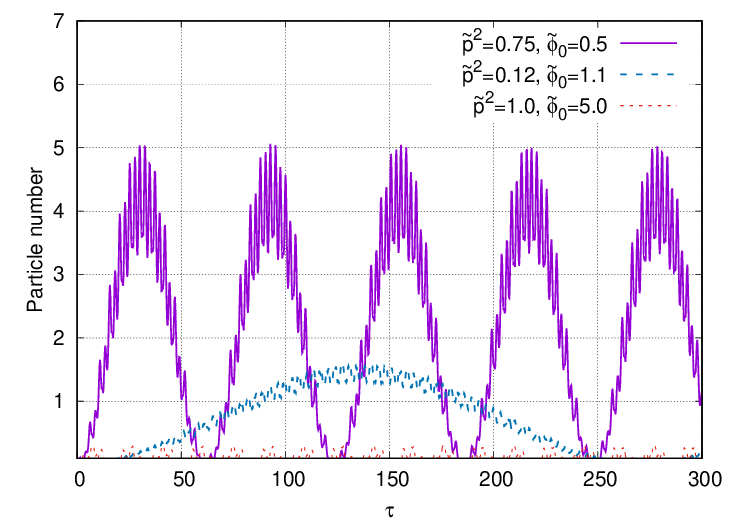}\\
(a) stable parameter
	\end{minipage}
	&
	\begin{minipage}[b]{0.55\linewidth}
		\centering
\includegraphics[scale=0.6]{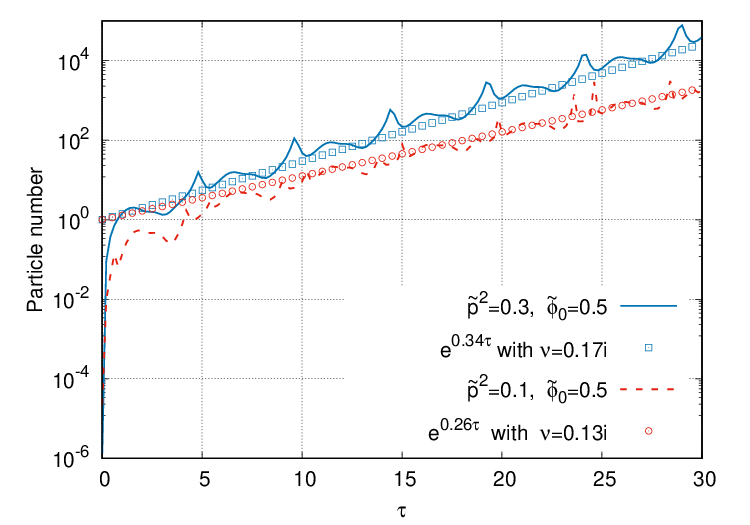}\\
(b) unstable parameter
	\end{minipage}
\end{tabular}
\caption{Particle numbers for stable/unstable parameters. \label{fig.particle.number.typical.stable.unstable}} 
\end{figure}

We plot the result of $\tilde{N}_p$ and $\bar{N}_p$ as the function of $\tau$ in Fig.~\ref{fig.particle.number.typical.stable.unstable}.
The plot (a) in Fig.~\ref{fig.particle.number.typical.stable.unstable} shows the particle production for the stable parameters region. The parameter which is located at near instability band give larger contribution to the particle production even in the stable region. On the other hand, the plot (b) in Fig.~\ref{fig.particle.number.typical.stable.unstable} shows the particle production for the unstable parameter region. As is expected from the exponential growth for the mode function in unstable region, namely, $\tilde{g}_p(\tau)\propto e^{\pm i\nu\tau}$ with a complex $\nu$, the particle production also shows the exponential growth, $e^{\pm 2i\nu\tau}$. We can expect that it will be possible to fit these exponential growths qualitatively by the Floquet's exponent for the unstable parameter region, namely, the complex Floquet's exponents $\nu=0.17i$ for the parametric amplification case $(\tilde{\phi}_0=0.3,\tilde{p}^2=0.5)$, and $\nu=0.13i$ for the spinodal instability case $(\tilde{\phi}_0=0.1,\tilde{p}^2=0.5)$. Actually this fit does not only qualitatively work well, but also fits well quantitatively, as is shown in the plot (b) of Fig.~\ref{fig.particle.number.typical.stable.unstable}.

\section{Discussion \label{sec4}}
We analyzed the mode equation based on the general property of the Hill's equation, namely, Floquet's theory. By numerical computations of the discriminant of the Hill's equation, $d$, we determined which parameter gives the stable/unstable behavior for the mode function. 
The instability parameter region is expressed by the band structure both in $(\tilde{\phi}_0, \tilde{p}^2)$-plane and $(\alpha, \tilde{p}^2)$-plane, as shown in Fig.~\ref{fig.discriminant.larger.than.2.p2.phi0.alpha.space}. 

Comparing the parametric amplification with the spinodal instability, it is more difficult to find the former than later, because later one can be estimated by the region which satisfies $\omega^2_p(t)<0$.  
Our analysis based on the general Hill's equation helps to understand the dynamics of the quantum mode around the nonlinear massive wave solution in the Higgs potential. As the consequence of the Floquet's theory, we obtained the Floquet's exponent which determines the later time dynamics of the quantum mode and the particle creation. The behavior of the particle creation in the unstable parameter region can be understood by using the results in Fig.~\ref{fig.Floquet.exp.p2.phi0.alpha.space}. 

An implication of our analysis will be for time dynamics of the Higgs phenomenon in early universe. The authors in Refs.~\cite{CosmoEffect1} studied the instabilities for the modified-Mathieu equation and the Whittaker-Hill equation to investigate the Higgs-inflaton and Higgs-gravity couplings. It will be possible to apply our analysis based on the nonlinear massive wave solution and Hill's equation to such analysis for quantum mode and argue how much the result will change. Because the nonlinear massive solution includes the nonlinear effect of the Higgs potential, we can investigate how much the classical nonlinear effect changes the research in literature. 

One of interesting remained analysis will be to obtain the instability band in Fig.~\ref{fig.discriminant.larger.than.2.p2.phi0.alpha.space} in terms of a compact analytical form. This question has the phenomenological importance for simplicity when we apply our result to other problems. The authors \cite{Herring.2024} obtained the approximated formula to express the instability band for their potential $+\mu^2\phi^2/2+\lambda \phi^4/4$ with the classical harmonic oscillation $\phi(t)=\phi(0)\cos(mt)$ by the series expansion of the parameters in the mode equation. In our case, our classical solution is described by Jacobi's elliptic function and we cannot use their result. It will be better for us to use mathematical theorems for the Hill's equation, for example, see Refs.~\cite{Hochstadt.1965,Loud.1975,Weinstein.1987,Mamode.2004,Keller.2007} for the case that the classical field has infinite terms in Fourier expansion and see Refs.~\cite{Levy.1963,Hochstadt.1964} for the case that the classical filed has a finite terms in Fourier expansion. 

Another important extension is to take into account effects of the cosmological expansion of the universe. In our analysis, the classical nonlinear oscillation in the Higgs potential is assumed to be periodical forever, namely, the amplitude does not decrease in time. However, if we apply our analysis to a realistic problem of cosmological particle productions, it is important to take into account the suppression effect of the scale factor in the expansion of the universe \cite{Turner.1983}. Then the classical field will be suppressed after a long time and thus the angular frequency will approach to a constant value, namely, the final state will have a fixed angular frequency. This effect will help to define the asymptotic final state for adiabatic particle. Moreover, the authors in Ref.~\cite{Herring.2024} proposed a conjecture that the system may have a new asymptotic state which minimizes the renormalized effective potential with the quantum correction as a consequence of the time evolution and the energy conservation of the system. If we can improve this point, we can investigate their conjecture for the system described by the quantum field theory coupling to the nonlinear massive wave solution in the Higgs potential. The authors in Ref.~\cite{Higgs.elliptic.ref1} discussed the asymptotic expansion method to discuss the nonlinear dynamics of the Higgs field with cosmological expansion. Their method will help us to improve our analysis.

\section{Conclusion \label{sec5}}
We discussed the quantum fluctuation around the nonlinear massive wave solution in the Higgs potential. In particular, we focus on the dynamics of the mode function in the quantum fluctuation and analyzed the instability/stability of the time evolution using the Floquet's theory for Hill's equation. We found the instability region as the function of parameters in the mode equation. As the consequence, we discussed the particle number for the zero-th-order adiabatic mode which is constructed by the linear combination of the creation/annihilation operators of the Higgs field with the unstable parameters. The behavior of the particle number can be understood by the imaginary part of the Floquet's exponent. 

The implication of our analysis to phenomenology is that the quantum field theory described in the article will be helpful to understand the time evolution of the system when the classical background solution of the EOM in the Higgs potential is slightly excited from VEV. This situation corresponds to the intermediate step of the phase transition of the electroweak theory. To improve our analysis will be possible by taking into account the suppression effect of nonlinear massive wave background with the scale factor or some approximations proposed in literature. Such improvement will enable us to test the proposed conjecture of the new equilibrium state which has not yet known. 

\bibliography{apssamp}

\end{document}